\documentclass[final,5p,times,twocolumn]{elsarticle}

\usepackage{graphicx}
\usepackage{amssymb}
 \usepackage{amsthm}

\usepackage{lineno}

\usepackage{subfigure}
\usepackage{booktabs}
\usepackage{verbatim}

\journal{NIM A  RICAP-2013}

\begin{document}

\begin{frontmatter}



\title{Preliminary results of ANAIS-25}


\author[uz,lsc]{J.~ Amar\'{e}}
\author[uz,lsc]{S.~Cebri\'{a}n}
\author[uz,lsc]{C.~Cuesta\corref{uw}\fnref{ca}}
\ead{ccuesta@uw.edu}
\cortext[ca]{Corresponding author}

\author[uz,lsc]{E.~Garc\'{\i}a}
\author[uz,lsc]{C.~Ginestra}
\author[uz,lsc,araid]{M.~Mart\'{\i}nez}
\author[uz,lsc]{M.~A. Oliv\'{a}n}
\author[uz,lsc]{Y.~Ortigoza}
\author[uz,lsc]{A.~Ortiz~de~Sol\'{o}rzano}
\author[uz,lsc]{C.~Pobes}
\author[uz,lsc]{J.~Puimed\'{o}n}
\author[uz,lsc]{M.~L.~Sarsa}
\author[uz,lsc]{J.~A.~Villar}
\author[uz,lsc]{P.~Villar}

 \address[uz]{Laboratorio de F\'{\i}sica Nuclear y Astropart\'{\i}culas, Universidad de Zaragoza, Calle Pedro Cerbuna 12, 50009 Zaragoza, Spain}
 \address[lsc]{Laboratorio Subterr\'{a}neo de Canfranc, Paseo de los Ayerbe s/n, 22880 Canfranc Estaci\'{o}n, Huesca, Spain}
\fntext[uw]{Presently at Center for Experimental Nuclear Physics and Astrophysics, and
Department of Physics, University of Washington, Seattle, WA, US}
\address[araid]{Fundaci\'{o}n ARAID, Mar\'{\i}a de Luna 11, Edificio CEEI Arag\'{o}n, 50018 Zaragoza, Spain}

\begin{abstract}
The ANAIS (Annual Modulation with NaI(Tl) Scintillators) experiment aims at the confirmation of the DAMA/LIBRA signal using the same target and technique at the Canfranc Underground Laboratory. 250\,kg of ultrapure NaI(Tl) crystals will be used as a target, divided into 20 modules, each coupled to two photomultipliers. Two NaI(Tl) crystals of 12.5\,kg each, grown by Alpha Spectra from a powder having a potassium level under the limit of our analytical techniques, form the ANAIS-25 set-up. The background contributions are being carefully studied and preliminary results are presented: their natural potassium content in the bulk has been quantified, as well as the uranium and thorium radioactive chains presence in the bulk through the discrimination of the corresponding alpha events by PSA, and due to the fast commissioning, the contribution from cosmogenic activated isotopes is clearly identified and their decay observed along the first months of data taking. Following the procedures established with ANAIS-0 and previous prototypes, bulk NaI(Tl) scintillation events selection and light collection efficiency have been also studied in ANAIS-25.
\end{abstract}

\begin{keyword}
Dark Matter \sep Annual modulation \sep Underground Physics \sep Sodium iodide scintillators


\end{keyword}

\end{frontmatter}

% \linenumbers

\section{Introduction}
\label{first}

The ANAIS (Annual modulation with NaI Scintillators) project is intended to search for dark matter annual modulation with 250\,kg of ultrapure NaI(Tl) scintillators at the Canfranc Underground Laboratory (LSC) in Spain. The motivation of the ANAIS experiment is to provide a model-independent confirmation of the annual modulation positive signal reported by the DAMA collaboration~\cite{dama} using the same target and technique. Several requisites have to be fulfilled: an energy threshold below 2\,keV in electron equivalent energy, a radioactive background lower than 2\,cpd/keV/kg in the lowest energy region, and very stable working conditions.

The total NaI(Tl) active mass will be divided in 20 modules, each consisting of a 12.5 kg NaI(Tl) crystal encapsulated in copper and optically coupled to two photomultipliers (PMTs) working in coincidence. The shielding foreseen for the experiment will consist of 10\,cm of archaeological lead, 20\,cm of low activity lead, 40\,cm of neutron moderator, an anti-radon box (to be continuously flushed with boil-off nitrogen), and an active muon veto system made up of plastic scintillators designed to cover top and sides of the whole ANAIS set-up. The hut that will house the experiment at the hall B of LSC (under 2450 m.w.e.), shielding materials and electronic chain components are prepared for mounting. Different PMT models have been tested in order to choose the best option in terms of light collection and background. The main challenge is still the achievement of the required low background, in particular the development of crystals having a potassium content at 20\,ppb level or below. Two prototypes of 12.5\,kg mass, made by Alpha Spectra with ultrapure NaI powder are taking data at the LSC since December 2012 for a general performance and background assessment. We will refer in the following to this set-up as ANAIS-25.

ANAIS-25 preliminary results are compared to those obtained with the former ANAIS-0 module~\cite{ANAISbkg,ANAIStaup11}, a parallelepipedic 9.6\,kg NaI(Tl) crystal with a size of 254.0\,x\,101.6\,x\,101.6\,mm$^{3}$, produced by Saint-Gobain, that was encapsulated using ETP copper at University of Zaragoza. The latter was operated at hall B of LSC from February, 2011 to December, 2012, inside a shielding consisting of 10\,cm archaeological lead plus 20\,cm low activity lead, all enclosed in a PVC box tightly closed and continuously flushed with boil-off nitrogen. Active vetoes were mounted only on top of the shielding to reject coincident events in ANAIS-0 module. Data were taken with different PMTs with and without light guides in order to characterize and fully understand ANAIS background at low and high energy, to optimize analysis and calibration methods, and to test the electronics and acquisition system.

\section{ANAIS-25 experimental set-up}
\label{second}

The ANAIS-25 set-up is formed by two cylindrical 12.5\,kg NaI(Tl) detectors grown with ultrapure NaI powder ($<$90\,ppb potassium at 95\% CL according to results from HPGe spectrometry analysis carried out at LSC) and built in collaboration with Alpha Spectra~\cite{AS}. The crystals were encapsulated in OFHC copper with two synthetic quartz windows allowing the PMTs coupling in a second step, as it was done with ANAIS-0. Only white Teflon was used as light diffuser, wrapping the crystal, inside the copper encapsulation. A Mylar window allows to calibrate at low energy both detectors.

\begin {figure}[ht]
\subfigure[]{\includegraphics[width=0.43\textwidth]{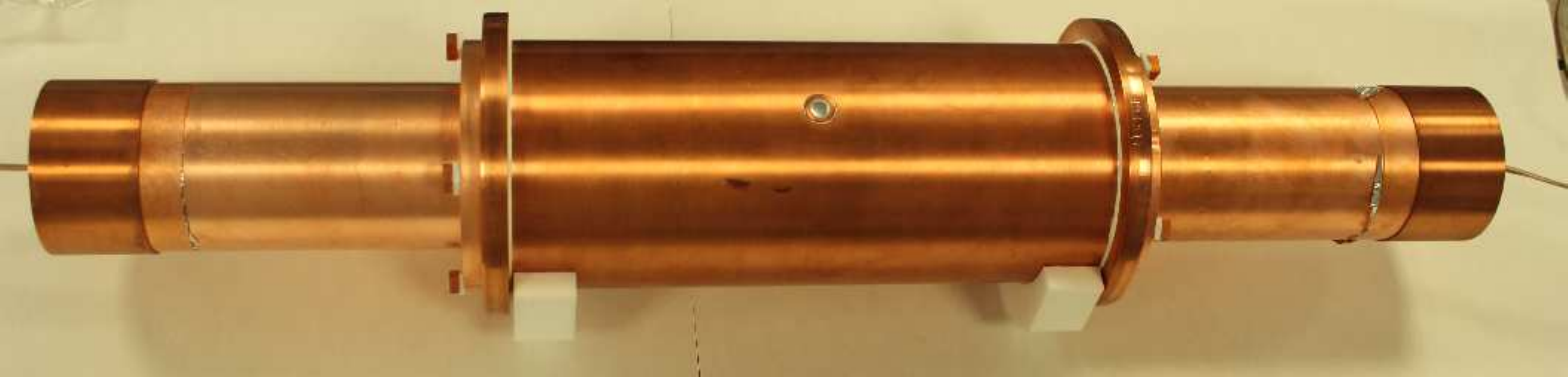}}
\subfigure[]{\includegraphics[width=0.4\textwidth]{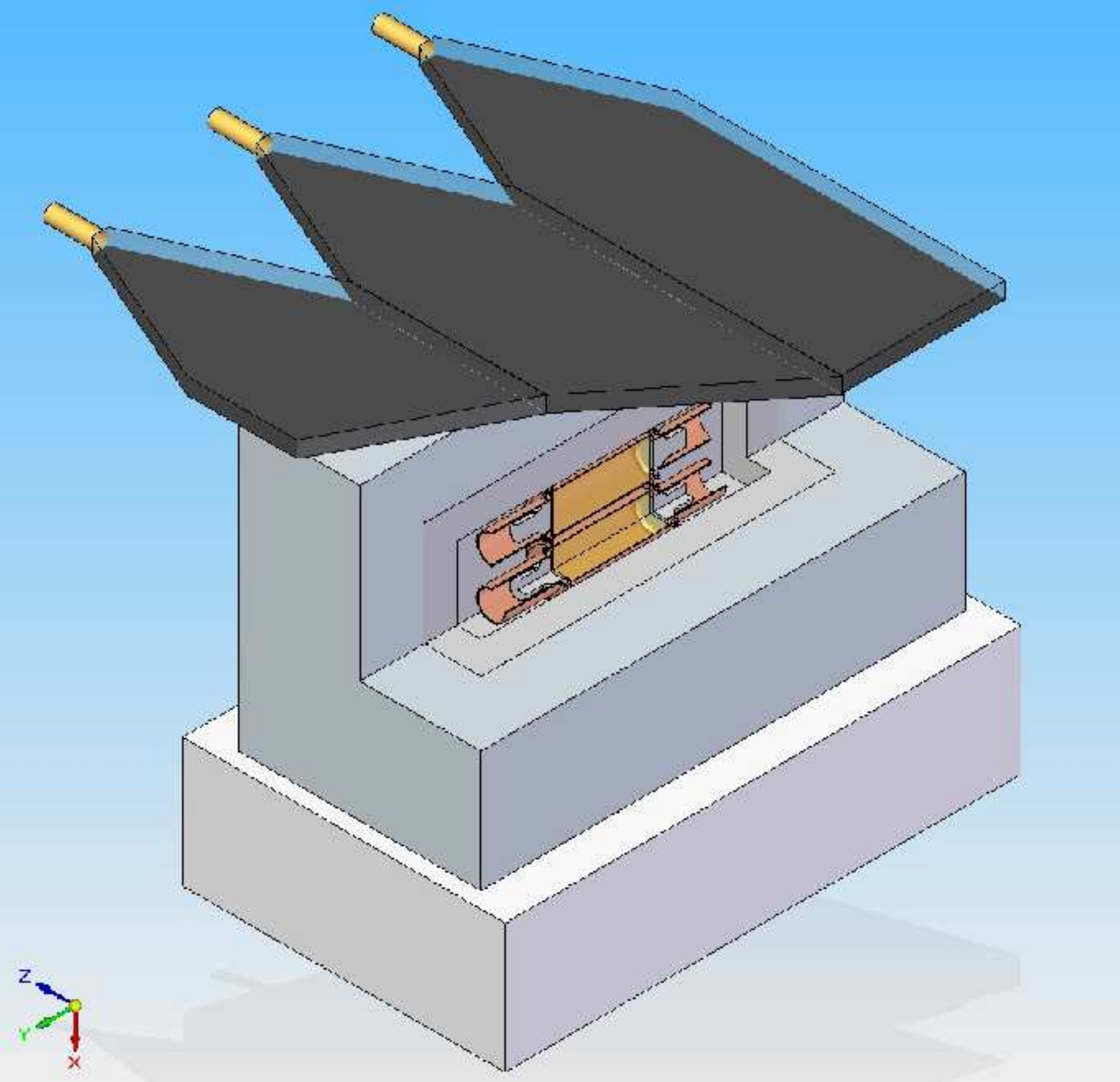}}
\centering \caption{\it (a) Final appearance of ANAIS-25 detector~0 after the coupling of the Ham. R12669SEL2 PMTs and placement of their copper casings at the LSC clean room. (b) Schematic drawing of the ANAIS-25 experimental layout at LSC consisting of 10\,cm archaeological lead plus 20\,cm low activity lead, all enclosed in a PVC box continuously flushed with boil-off nitrogen and active vetoes anti-muons.}
\label{fig:A25_d}
\end {figure}

They were shipped by boat from the US and arrived at LSC in December 2012 and data taking started only a few days after moving them underground and coupling two PMTs to each detector in the LSC clean room. Hamamatsu (Ham.) R12669SEL2 PMTs were used for one crystal (detector~0), shown in Fig.~\ref{fig:A25_d}.a, and Ham. R11065SEL PMTs for the other (detector~1). ANAIS-25 experimental layout can be seen in Fig.~\ref{fig:A25_d}.b. It has been operated in very similar experimental conditions than the previous ANAIS-0 prototype~\cite{ANAISbkg}.

\section{Background understanding}
\label{fourth}

Main goal of ANAIS-25 set-up was to determine the potassium content of the crystals by the coincidence technique. At the same time, $^{238}$U and $^{232}$Th chains isotopes content in the crystals had to be quantified, as well as total background of the two modules be assessed. %As the modules started to take data just a couple of days after their arrival, cosmogenic activation was clearly observable in the first weeks of data.

\begin {figure}[ht]
\includegraphics[width=0.42\textwidth]{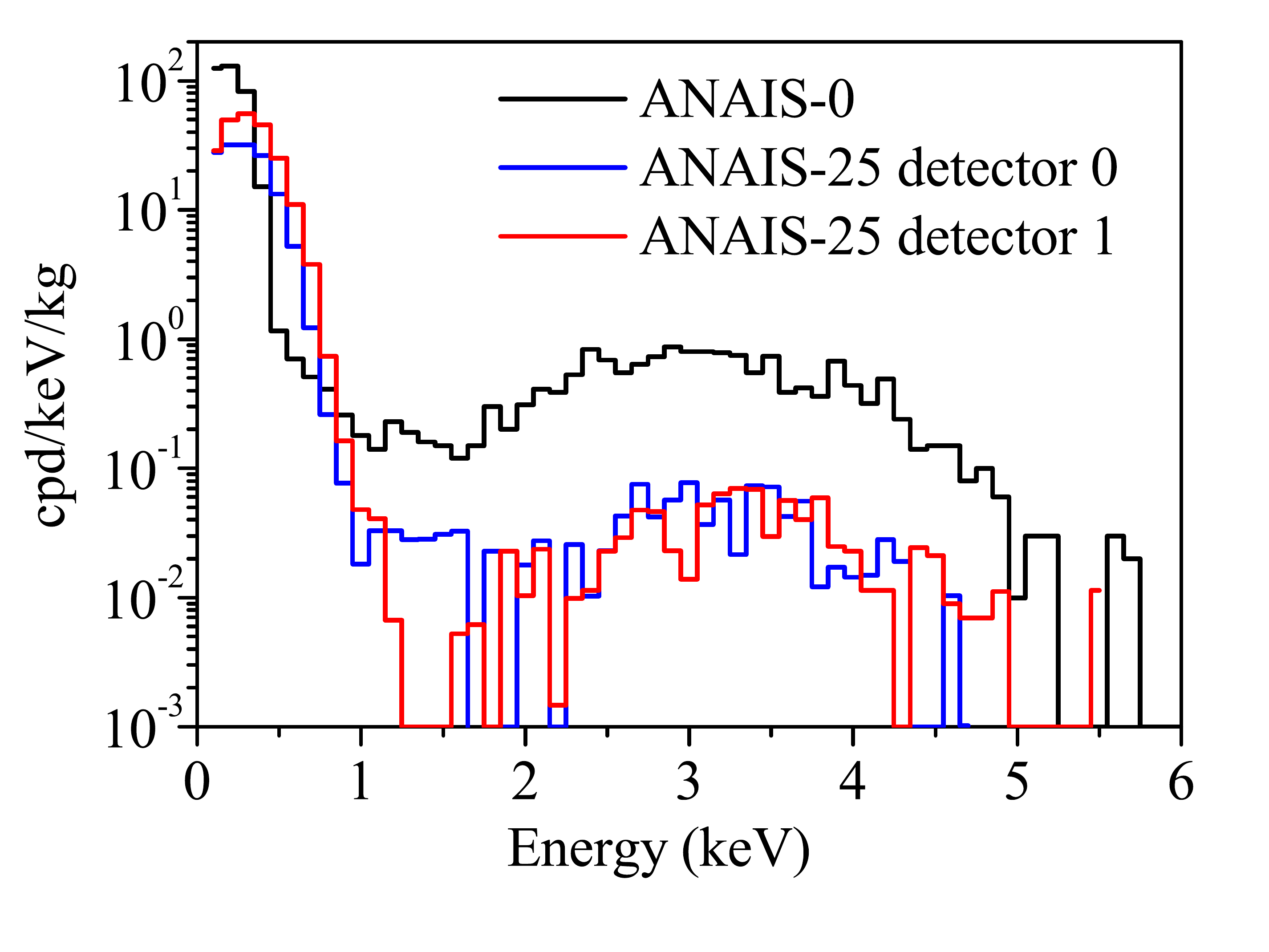}
\centering \caption{\it Low energy spectra in coincidence with 1\,$\sigma$ windows around 1461\,keV line in the other crystal for ANAIS-0 (black), ANAIS-25 detector~0 (blue), and ANAIS-25 detector~1 (red).}
\label{fig:A25_40K_LEcn}
\end {figure}

The potassium content of the ANAIS-25 crystals has been carefully analyzed using the same technique applied to previous prototypes. Bulk $^{40}$K content is estimated by searching for the coincidences between 3.2\,keV energy deposition in one detector (following EC) and the 1460.8\,keV gamma line escaping from it and being fully absorbed in the other detector. Efficiency of the coincidence was estimated using Geant4. Good agreement between results derived for both detectors is observed. We can conclude that ANAIS-25 crystals have a $^{40}$K content of 1.25$\pm$0.11\,mBq/kg (41.7$\pm$3.7\,ppb of potassium), one order of magnitude better than that found in ANAIS-0 crystal (see Fig.~\ref{fig:A25_40K_LEcn}). However, the 20\,ppb goal has not been achieved and before ordering the additional 18 modules required to complete the ANAIS total detector mass, careful analysis of the situation in collaboration with Alpha Spectra is required.

Activities of the different branches in $^{238}$U and $^{232}$Th chains could be precisely identified in ANAIS-0 prototype by their alpha emissions, discriminated by Pulse Shape Analysis (PSA). The total alpha rate measured in ANAIS-25 is 3.15\,mBq/kg, much higher than that of ANAIS-0. The $^{232}$Th natural chain seems to be really suppressed in ANAIS-25 crystals, as points out the very low number of $^{212}$Bi-Po coincidences identified. Hence, we have to attribute such a rate to isotopes from the $^{238}$U chain, probably out of equilibrium, because we do not see the expected alpha lines structure. The presence of $^{210}$Pb events in the low energy range, confirms the assumption that most of the alpha events observed could be coming from this part of $^{238}$U chain. More statistics is required in order to properly calibrate the alpha spectrum and to determine the precise contribution from each component.

\begin {figure}[ht]
\subfigure[]{\includegraphics[width=0.41\textwidth]{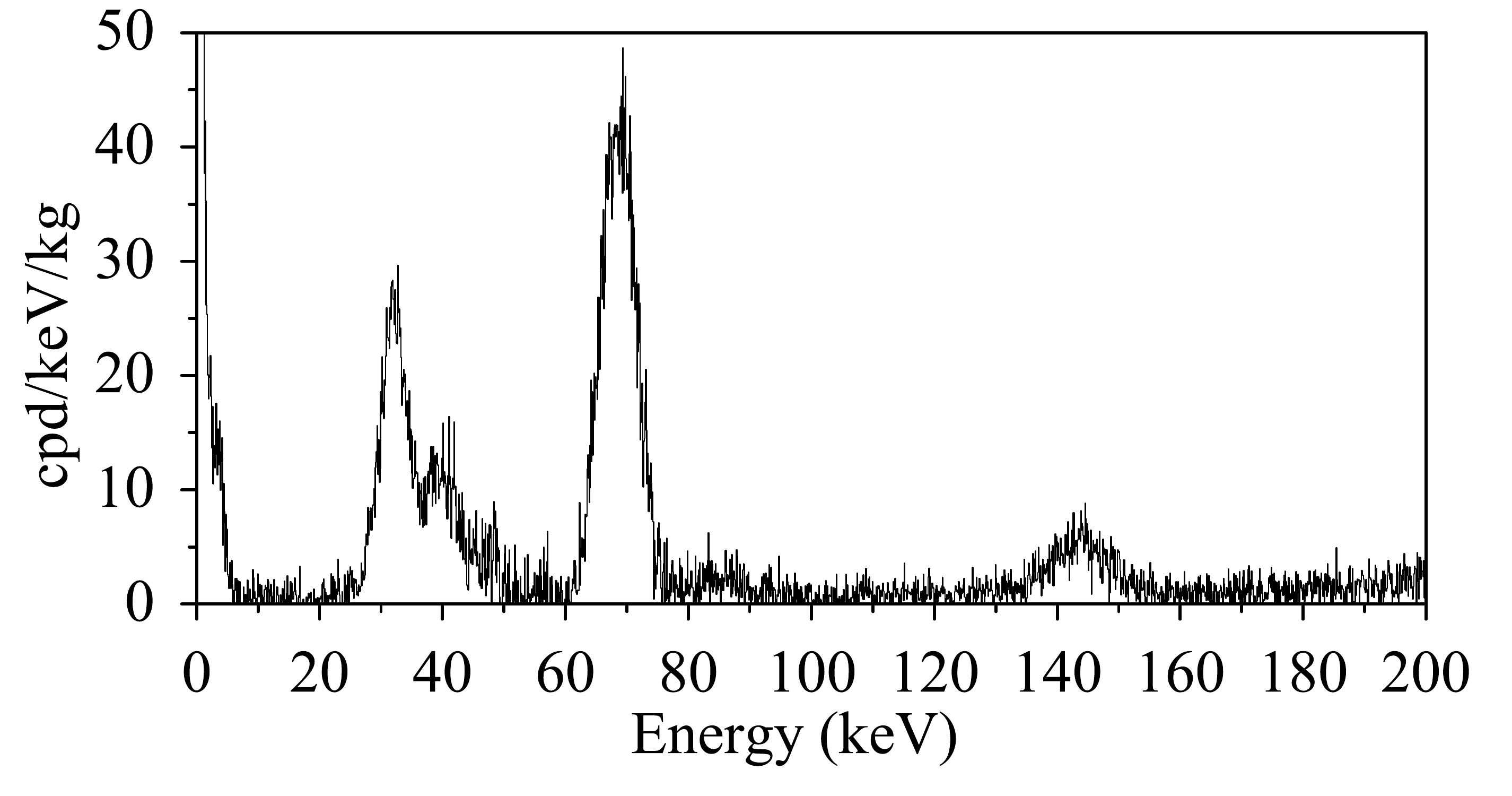}}
\subfigure[]{\includegraphics[width=0.41\textwidth]{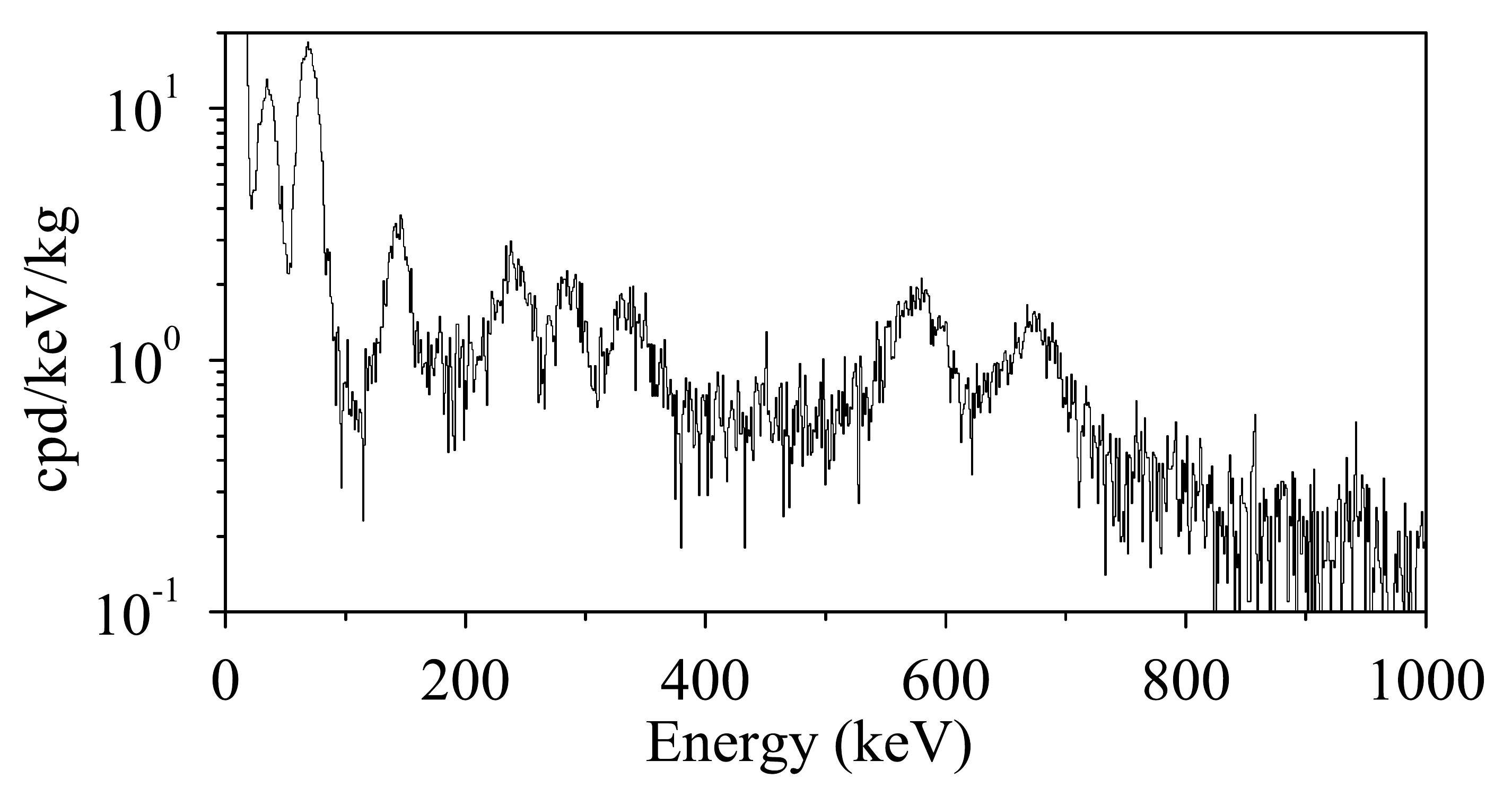}}
\centering \caption{\it Difference of the spectra corresponding to the first week ANAIS-25 detector~0 data underground and data taken a week starting 75 days later. (a) In the low energy region, emissions from $^{125}$I (35.5\,keV plus emissions following L and K electron capture in Te) at 40.4 and 67.3\,keV, $^{127m}$Te at 88.3\,keV and $^{125m}$Te at 144.8\,keV have been identified, while peaks around 4-5\,keV and 30-32\,keV are compatible with emissions following L and K electron capture in Te/I isotopes. (b) In the high energy range, emissions from $^{123m}$Te at 247.6\,keV, $^{121m}$Te at 294.0\,keV, $^{121}$Te at 507.6 and 573.1\,keV and $^{126}$I at 666.0\,keV have been identified.}
\label{fig:A25_cosmo}
\end {figure}

ANAIS-25 detectors started to take data just three days after going underground. This allowed to observe short-life isotopes activated during the stay on surface of all detectors components, mainly the NaI crystals. Besides the prompt data taking starting at LSC, low radioactivity level of the modules and very good resolution have contributed significantly to that issue. Fig.~\ref{fig:A25_cosmo} shows the spectra in the high and low energy regions corresponding to the difference between first week of data and those obtained a week starting 75 days after. Several lines are clearly attributable to cosmogenic activation in the subtracted spectra. Main isotopes identified are shown in Table~\ref{tab:cosmoiso}. Radon contamination in the inner shielding cavity was present in the first weeks of measurements, and some of the lines observed in Fig.~\ref{fig:A25_cosmo} have such origin.

\begin{table}[ht]
\begin{center}
\fboxrule=0cm  \fbox{
\begin{tabular}[]{llll}
\toprule
Isotope		& Lifetime	& Decay					& Main emissions \\
\cmidrule{2-4}

			& days 		& 			            & keV \\
\cmidrule{1-4}
$^{125}$I 	& 59.4 		& EC					& 35.5\\
$^{126}$I 	& 13.11		& EC, $\beta^{-}$		& 666.0\\
$^{121m}$Te & 154 		& IT, EC				& 294.0\\
$^{121}$Te 	& 16.8		& EC					& 507.6, 573.1\\
$^{123m}$Te & 119.7		& IT       				& 247.6\\
$^{125m}$Te & 57.4 		& IT 					& 144.8\\
$^{127m}$Te & 109		& IT, $\beta^{-}$		& 88.3\\
\bottomrule
\end{tabular}
}\caption{\it Main cosmogenic isotopes identified in the first weeks of data with the ANAIS-25 detectors underground. Their lifetime, decay, and main emission are shown. Data obtained from~\cite{nucleide,isotopes}.}
\label{tab:cosmoiso}
\end{center}
\end{table}

\begin {figure}[ht]
\includegraphics[width=0.41\textwidth]{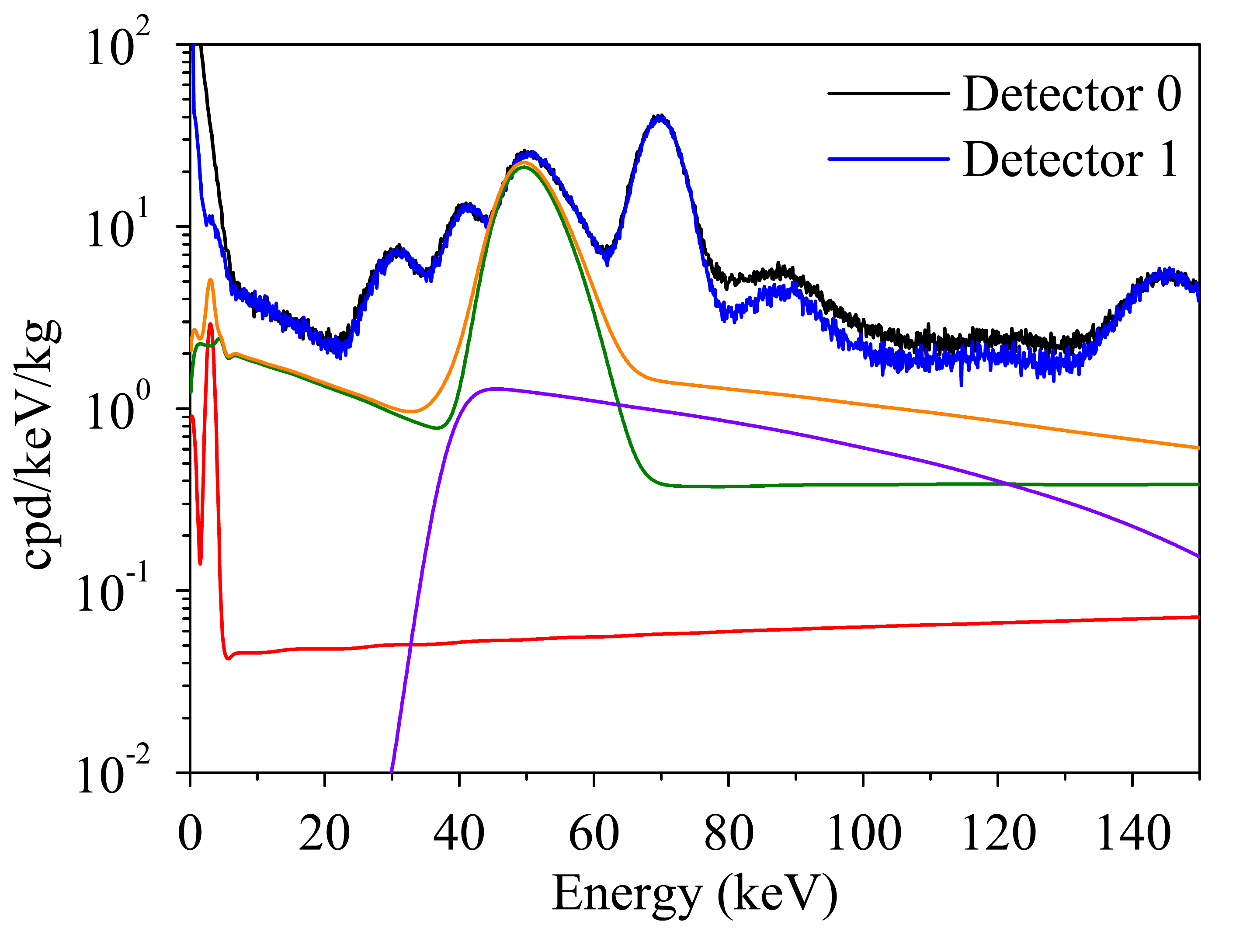}
\centering \caption{\it Total background spectra of ANAIS-25 detector~0 (black) and 1 (blue), only raw data are shown. Simulated contamination of 1.25\,mBq/kg of $^{40}$K (red), 3.15\,mBq/kg of $^{210}$Pb (green), and 0.94\,mBq/kg of $^{129}$I (violet) are also shown. Cosmogenic isotopes contributing to the spectra have not been simulated. The total of the simulated contaminations is shown in orange.}
\label{fig:A25_sim}
\end {figure}

The same simulation code developed for ANAIS-0~\cite{ANAISbkg} has been extended to the ANAIS-25 set-up by modifying accordingly the geometry. For the moment, only NaI bulk crystal contaminations have been taken into account and no simulation of the cosmogenically produced isotopes has been attempted. In Fig.~\ref{fig:A25_sim} the contribution to the background derived from the simulation for 1.25\,mBq/kg of $^{40}$K, 3.15\,mBq/kg of $^{210}$Pb (corresponding to the total alpha rate), and 0.94\,mBq/kg of $^{129}$I (the same specific activity used for ANAIS-0 in the background model presented in~\cite{ANAISbkg}) homogeneously distributed in the NaI(Tl) crystal are shown together with the low energy background spectra of ANAIS-25 detectors in the first months of measurement. It can be concluded that main features of the low energy background can be explained by a $^{210}$Pb contamination in the bulk.

\section{Light collection results}
\label{third}

Single photoelectron response (S.E.R.) has been obtained by identifying the last peak in the pulse for the four PMTs. To study the light collection efficiency of the ANAIS-25 detectors, the number of photoelectrons per keV (phe./keV) has been estimated from the S.E.R. area distribution and the area distribution of events corresponding to 22.6\,keV line from $^{109}$Cd calibration data. Results for the two detectors, and those obtained with ANAIS-0 using the same PMTs, are shown in Table~\ref{tab:ASphekev}.

\begin{table}[ht]
\begin{center}
\fboxrule=0cm \fboxsep=0cm \fbox{
\begin{tabular}[]{lll}
\toprule
PMT model      & ANAIS-0 	& ANAIS-25									\\
 \cmidrule{2-3}
                &  phe./keV	& phe./keV 											\\
 \cmidrule{1-3}
Ham. R12669SEL2 		& 7.38$\pm$0.07&16.13$\pm$0.66\\
Ham. R11065SEL		   & 5.34$\pm$0.05&12.58$\pm$0.13\\
\bottomrule
\end{tabular}
}\caption{\it Light collection efficiencies (phe./keV) for ANAIS-0 and ANAIS-25 detectors, derived from the 22.6\,keV line ($^{109}$Cd calibration). Excellent light collection efficiencies have been determined for ANAIS-25 modules.}
\label{tab:ASphekev}
\end{center}
\end{table}

\begin {figure}[ht]
\includegraphics[width=0.43\textwidth]{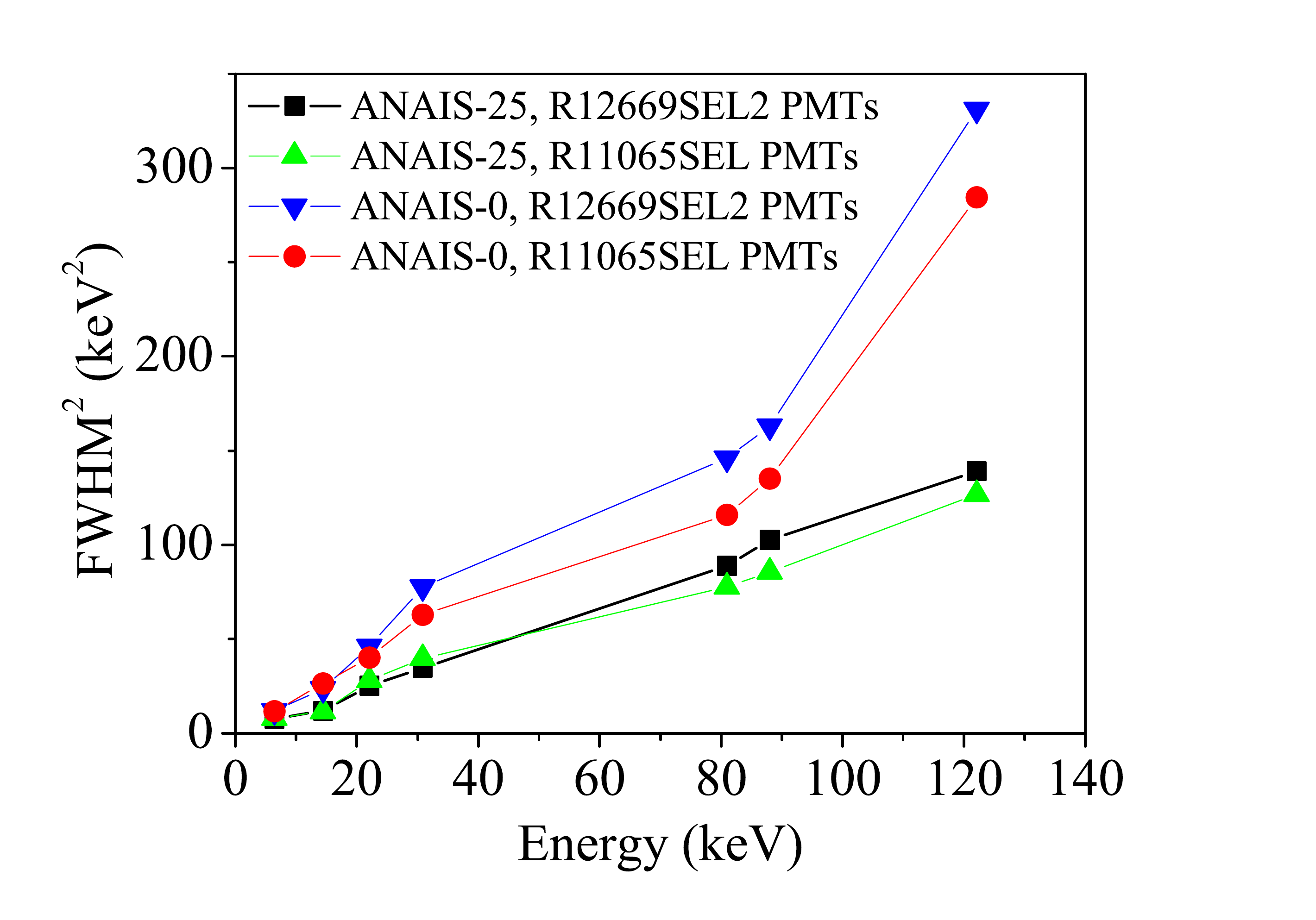}
\centering \caption{\it FWHM$^{2}$ for the different gamma lines measured at LSC in ANAIS-25 detectors compared with the ANAIS-0 module results when using the same PMTs.}
\label{fig:ASFWHM}
\end {figure}

Energy resolution of the ANAIS-25 detectors has been studied using data from $^{57}$Co, $^{133}$Ba and $^{109}$Cd calibrations. Results for the FWHM of the ANAIS-25 detectors are shown in Fig.~\ref{fig:ASFWHM}, compared to that of the ANAIS-0 module.

The obtained light collection efficiencies and resolution results are consistent with the better quantum efficiency of the Ham. R12669SEL2 PMTs, used in detector~0 from ANAIS-25 set-up, already stated with ANAIS-0 data. Moreover, the light collection efficiencies of the ANAIS-25 detectors are much better than those previously obtained with ANAIS-0. This remarkable improvement is attributed to the much better optical coupling of the new modules and the excellent optical quality of the NaI(Tl) crystals.

\section{Events selection}
\label{fifth}

\begin {figure}[ht]
\includegraphics[width=0.45\textwidth]{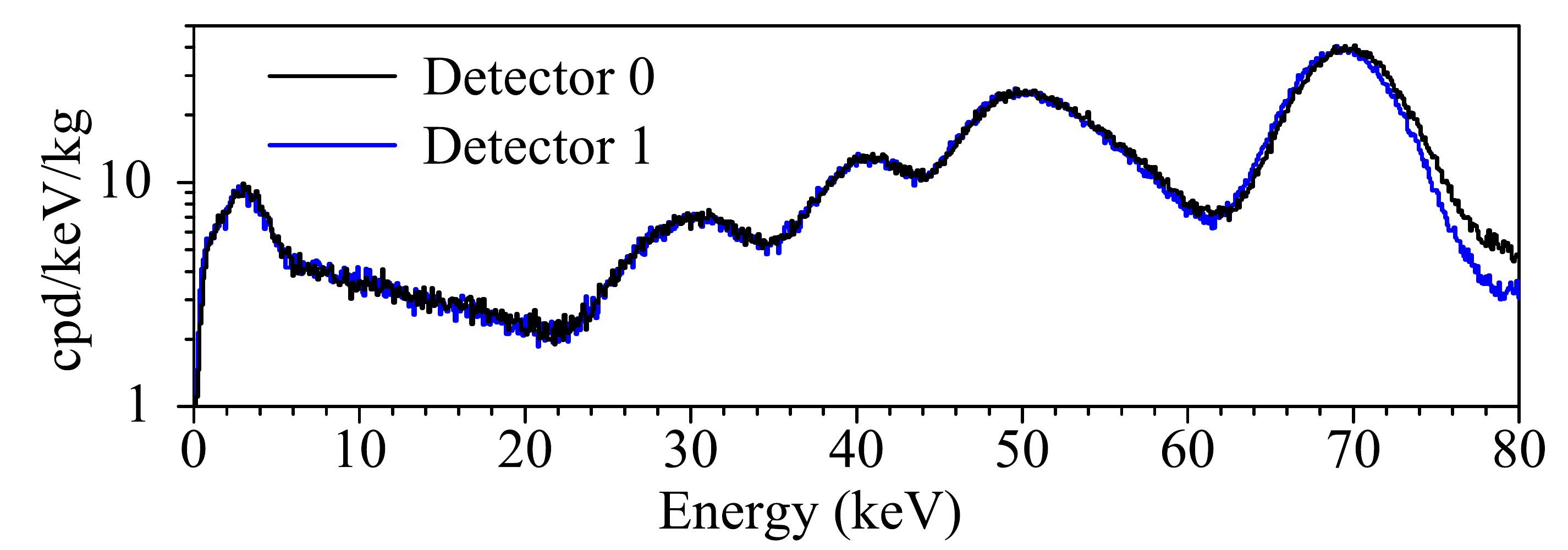}
\centering \caption{\it Low energy spectra of ANAIS-25 detectors after having applied all the cuts described in text.}
\label{fig:A25LEcut}
\end {figure}

Rejection of non bulk NaI scintillation events is required to reduce the effective threshold, because event rate below 10\,keV is dominated by non bulk NaI scintillation events, specially in detector~0 (see Fig.~\ref{fig:A25_sim}). The events selection protocol for ANAIS-25 data is not completed yet, but a preliminary filtering procedure has been applied, following that developed for ANAIS-0. Scarce events with anomalous baseline estimate have been rejected. As in ANAIS-25 set-up two detectors are taking data, an anticoincidence cut has been also implemented. At last, the cut on the peaks number of the event (related to the number of discrete  photoelectrons identified in the pulse) has been used. For the detector~1 the same criterion as in ANAIS-0 has been followed and events having less than 3 peaks in any of the PMT signals have been discarded. As detector~0 presents a higher dark rate, events having less than 5 peaks in any of the PMTs are rejected. In both detectors this cut implies an effective analysis threshold below 1\,keV, because of the excellent light collection efficiency of the ANAIS-25 detectors. In Fig.~\ref{fig:A25LEcut} the filtered spectra of both ANAIS-25 detectors are shown.

After the explained and preliminary filtering procedure applied, both detectors present the same background, pointing at a negligible contribution of the PMTs to the low energy background, in spite of their different radioactivity levels~\cite{ANAISbkg}. If this point is confirmed after the decay of cosmogenic isotopes, light guides would be avoided in ANAIS.

\appendix

\section*{Summary}

First background data of two new 12.5\,kg detectors forming ANAIS-25 have been analyzed, showing a K content of 41.7$\pm$3.7\,ppb. Cosmogenic activation in NaI is under study: several short-life isotopes have been clearly identified. After substantial decay of cosmogenic isotopes, a thorough understanding of background contributions is being pursued, in collaboration with Alpha Spectra, for a general background assessment of the ANAIS-25 set-up focusing in the $^{210}$Pb content. An excellent light collection has been measured in both ANAIS-25 detectors (12-16\,phe./keV). On view of the good results derived using the Ham. R12669SEL2 PMT model, it has been chosen to be used in ANAIS and 42 units have been ordered.

\section*{Acknowledgements}

This work has been supported by the Spanish Ministerio de Econom\'{\i}a y Competitividad and the European Regional Development Fund (MINECO-FEDER) (FPA2011-23749), the Consolider-Ingenio 2010 Programme under grants MULTIDARK CSD2009- 00064 and CPAN CSD2007-00042, and the Gobierno de Arag\'{o}n (Group in Nuclear and Astroparticle Physics, ARAID Foundation and C. Cuesta predoctoral grant). C. Ginestra and P. Villar are supported by the MINECO Subprograma de Formaci\'{o}n de Personal Investigador. We also acknowledge LSC and GIFNA staff for their support.

\section*{References}
%% References


\bibliographystyle{elsarticle-num}
\bibliography{ANAIS25_Ricap13}






% \end{thebibliography}

\end{document}